\documentclass[aps,prl,
 amsmath,amssymb,
 twocolumn, groupedaddress,
 showpacs]{revtex4-1}
\usepackage{bm}
\usepackage{graphicx}
\usepackage{gensymb}
\usepackage{amsmath}
\usepackage{amssymb}
\usepackage{color}
\usepackage{ulem}

\begin{document}

\title{Monolayer VTe$_2$: incommensurate Fermi-surface nesting and suppression of charge density wave}

\author{Katsuaki Sugawara,$^{1,2,3}$ Yuki Nakata,$^{1}$ Kazuki Fujii,$^{1}$ Kosuke Nakayama,$^{1}$ Seigo Souma,$^{2,3}$ Takashi Takahashi,$^{1,2,3}$ and Takafumi Sato$^{1,2,3}$}

\affiliation{$^1$Department of Physics, Tohoku University, Sendai 980-8578, Japan}
\affiliation{$^2$Center for Spintronics Research Network, Tohoku University, Sendai 980-8577, Japan}
\affiliation{$^3$WPI Research Center, Advanced Institute for Materials Research, Tohoku University, Sendai 980-8577, Japan}

\date{\today}

\begin{abstract}
We investigated the electronic structure of monolayer VTe$_2$ grown on bilayer graphene by angle-resolved photoemission spectroscopy (ARPES). We found that monolayer VTe$_2$ takes the octahedral 1$T$ structure in contrast to the monoclinic one in the bulk, as evidenced by the good agreement in the Fermi-surface topology between ARPES results and first-principles band calculations for octahedral monolayer 1$T$-VTe$_2$.  We have revealed that monolayer 1$T$-VTe$_2$ at low temperature is characterized by a metallic state whereas the nesting condition is better than that of isostructural monolayer VSe$_2$ which undergoes a CDW transition to insulator at low temperature. The present result suggests an importance of Fermi-surface topology for characterizing the CDW properties of monolayer TMDs.
\end{abstract}

\maketitle
 Layered transition-metal dichalcogenides (TMDs) are a promising candidate for realizing outstanding properties associated with two-dimensionalization since bulk TMDs are known to exhibit various physical properties such as magnetism, Mott-insulator phase, and charge density wave (CDW), besides the wide range of transport property (insulator, semiconductor, metal, and superconductor), most of which are prone to the change in the dimensionality of materials. When TMDs are thinned to a single monolayer (2D limit), they exhibit even more outstanding properties distinct from bulk, as represented by the room-temperature ferromagnetism in VSe$_2$ in contrast to the nonmagnetic nature of bulk \cite{Bonilla} and the change in the band-gap property from indirect to direct transition in MoS$_2$ \cite{Mak3}. The CDW is a most pronounced phenomenon widely seen in both bulk and atomic-layer TMDs. In bulk TMDs, the interplay between the Fermi-surface nesting and the energy-gap opening as well as its relationship to the strength of CDW properties such as the CDW transition temperature ($T_{\rm CDW}$) has been a target of intensive studies \cite{Rossnagel}.

The role of dimensionality to the mechanism of CDW, in particular whether or not the CDW is more stable in the 2D limit, is now becoming a target of fierce debates, being stimulated by a recent success in fabricating various atomic-layer TMDs by exfoliation and epitaxial techniques.  Recent studies on some TMDs such as TiSe$_2$, VSe$_2$, and NbSe$_2$ \cite{Chen, Sugawara, Umemoto, Feng, Xi} have shown a marked increase in $T_{\rm CDW}$ upon reducing the thickness down to a few monolayers. In contrast,  it has been reported that the CDW vanishes in monolayer TaS$_2$ and TaSe$_2$ \cite{Sanders, Shao}, suggesting an important role of substrate and many-body effects. In 1$T$-VSe$_2$, reducing the number of layers by exfoliating bulk crystal leads at first to gradual decrease of $T_{\rm CDW}$, but at a critical film thickness of $\sim$ 10 nm, the $T_{\rm CDW}$ exhibits a characteristic upturn and reaches 140 K in a few monolayers, much higher than bulk $T_{\rm CDW}$ (110 K) \cite{Yang}. Such enhancement of $T_{\rm CDW}$ in monolayer VSe$_2$ was also revealed by angle-resolved photoemission spectroscopy (ARPES) wherein the role of FS nesting and electron-phonon coupling was intensively debated \cite{Umemoto, Chen, Chen2, Feng, Duvjir, Zhang}. However, the nature of 2D CDW is still far from reaching a consensus, as highlighted by a wide variety of periodic lattice distortions hitherto proposed for monolayer VSe$_2$ (e.g., 4 $\times$ 4, 4 $\times$ 1, $\sqrt{7}$ $\times$ $\sqrt{3}$, 4 $\times$ $\sqrt{3}$) \cite{Umemoto, Chen2, Feng, Duvjir, Zhang}. In a broader perspective, it is still unclear to what extent the conventional FS-nesting picture can be applied to the 2D CDW materials and how the FS topology and electronic interactions are related to the CDW properties such as the enhancement/suppression of $T_{\rm CDW}$. To address these essential questions, a further study on the electronic structure with new monolayer TMDs is highly required.

In this Rapid Communication, we report a successful fabrication of monolayer VTe$_2$ on bilayer graphene / SiC and its ARPES investigation. While the monoclinic (1$T$'') phase is known to be stable in bulk VTe$_2$, our monolayer VTe$_2$ film takes the octahedral 1$T$ structure [see Fig. 1(a)]. Importantly, this enables us to directly compare the electronic states and CDW properties with isostructural monolayer 1$T$-VSe$_2$. Our ARPES study on monolayer VTe$_2$ signifies a large, nearly perfectly nested triangular FS centered at the K point in the Brillouin zone (BZ).  We also found that the V 3$d$ band apparently crosses the Fermi level ($E_{\rm F}$) midway between the $\Gamma$ and K points, indicative of the metallic state at low temperature, unlike monolayer VSe$_2$ which shows a fully gaped insulating state below 140 K. We discuss possible origins for such an intriguing difference in terms of the variation in the FS-nesting condition and electron-phonon coupling.

A high-quality monolayer VTe$_2$ film was grown on bilayer graphene by the molecular-beam-epitaxy (MBE) method. ARPES measurements were performed at the BL-28B beamline in Photon Factory, KEK. First-principles band-structure calculations were carried out using the QUANTUM ESPRESSO code with generalized gradient approximation \cite{QunatumEsp, GGA}. For details, see section 1 of Supplemental Material \cite{SM}.

\begin{figure}
\begin{center}
\includegraphics[width=3.0in]{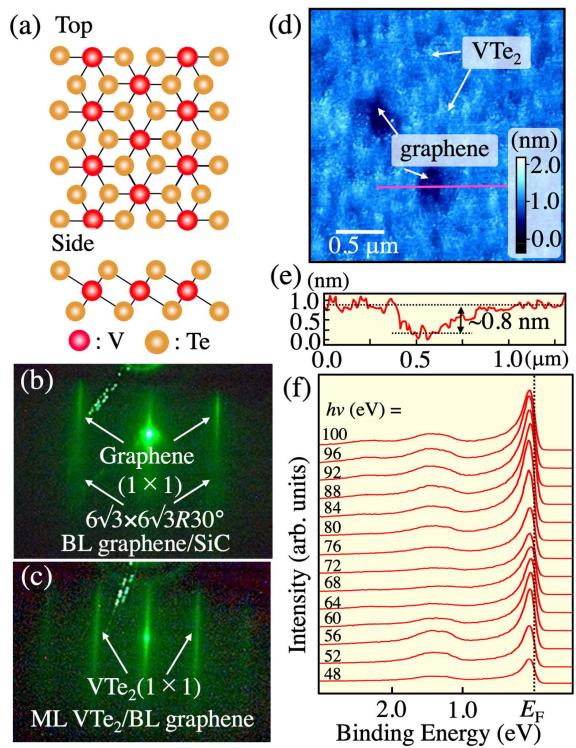}
\caption{(a) Crystal structure of monolayer 1$T$-VTe$_2$. (b),(c) RHEED patterns of bilayer (BL) graphene and monolayer (ML) VTe$_2$ on BL graphene, respectively. (d) AFM image of monolayer VTe$_2$. High (white), medium (blue), and low (dark) intensity regions are attributed to atoms/molecules adsorbed on VTe$_2$, clean monolayer VTe$_2$, and atoms molecules adsorbed on graphene substrate, respectively. The reason why we did not attribute the dominant white and blue areas to graphene or 2 monolayer VTe$_2$ is because we performed $in$-$situ$ ARPES measurements with the same sample and observed a dominant contribution from the monolayer energy bands to the total ARPES intensity. (e) Height profile along a cut shown by magenta solid line in (d). (f) Photon-energy dependence of the EDC at the $\Gamma$ point in monolayer VTe$_2$.}
 \end{center}
\end{figure}

First, we present characterization of monolayer VTe$_2$ film. Figure 1(b) shows the reflection high-energy electron diffraction (RHEED) pattern of pristine bilayer graphene on SiC(0001) substrate. We clearly observe the 1 $\times$ 1 and 6$\sqrt{3}$ $\times$ 6$\sqrt{3}$ $R$30$^\circ$ streak patterns, which correspond to bilayer graphene and underlying carbon-mesh layer on SiC, respectively \cite{Emtsev}. After growing a VTe$_2$ film by co-depositing V and Te atoms onto the bilayer-graphene surface, the RHEED intensity from the substrate disappears, and a sharp 1 $\times$ 1 streak pattern appears [Fig. 1(c)], similarly to the case of other monolayer TMD films grown on bilayer graphene \cite{Sugawara2, Nakata, Umemoto}, indicating the formation of VTe$_2$. Absence of additional streak patterns suggests no inclusion of monoclinic phase with zigzag chain structure which is known to exist in bulk VTe$_2$ below 482 K \cite{Bronsema, Ohtani}. This situation is particularly suited for comparing the electronic states with other 1$T$ monolayer polymorphs. We come back to this point later in detail. As seen in the $ex$-$situ$ atomic force microscopy (AFM) image in Fig. 1(d), large islands with a typical height of 0.8 nm, which corresponds to that of monolayer [Fig. 1(e)], are recognized on bilayer graphene. The energy distribution curve (EDC) at the $\Gamma$ point in Fig. 1(f) signifies no detectable photon-energy ($h\nu$) variation in the energy position of bands, supporting the 2D nature of electronic states.

We have estimated the in-plane lattice constant of monolayer VTe$_2$ as $a$ $\sim$ 3.35 $\rm{\AA}$ at room temperature by comparing the relative position of the RHEED patterns between graphene and VTe$_2$. This value is in good agreement with that estimated from the absolute wave vector values at the M (1.08 $\pm$ 0.03 \AA$^{-1}$) and K (1.24 $\pm$ 0.03 \AA$^{-1}$) points relative to the $\Gamma$ point in the ARPES data (3.35 $\pm$ 0.09 \AA). (including error bars due to the angular resolution and the angle-to-${\bm k}$ conversion). Intriguingly, these values are much smaller than that of bulk octahedral 1$T$-VTe$_2$ (3.64 \AA; obtained above 482 K where the 1$T$ phase is stable). Taking into account that the coupling between graphene and 1$T$-VTe$_2$ film could be sufficiently weak due to the van-der-Waals-coupling nature, it is inferred that the value of 3.35 \AA $ $  could be the most stable lattice parameter for free-standing monolayer VTe$_2$.

\begin{figure}
\begin{center}
\includegraphics[width=3.0in]{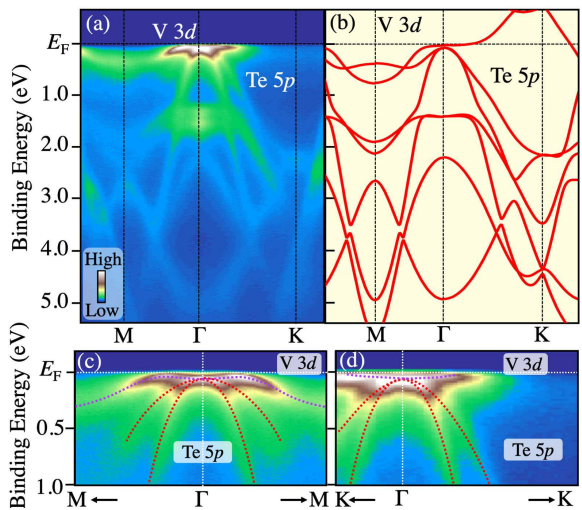}
\caption{(a) Plot of ARPES intensity for monolayer VTe$_2$ along the $\Gamma$M and $\Gamma$K cuts measured with $h\nu$ = 56 eV at $T$ = 40 K. (b) Band structure obtained from the first-principles band-structure calculations for monolayer 1$T$-VTe$_2$ with the input of experimental in-plane lattice constant (3.35 \AA). Overall calculated bands were contracted by 13 $\%$ in the energy axis to find a reasonable matching with the experiment. (c),(d) Experimental band structure near $E_{\rm F}$, measured along the $\Gamma$M and $\Gamma$K cuts, respectively. Red and purple dashed curves are a guide for the eyes to trace the Te 5$p$ and V 3$d$ bands, respectively.}
 \end{center}
\end{figure}

\begin{figure}
\begin{center}
\includegraphics[width=3.0in]{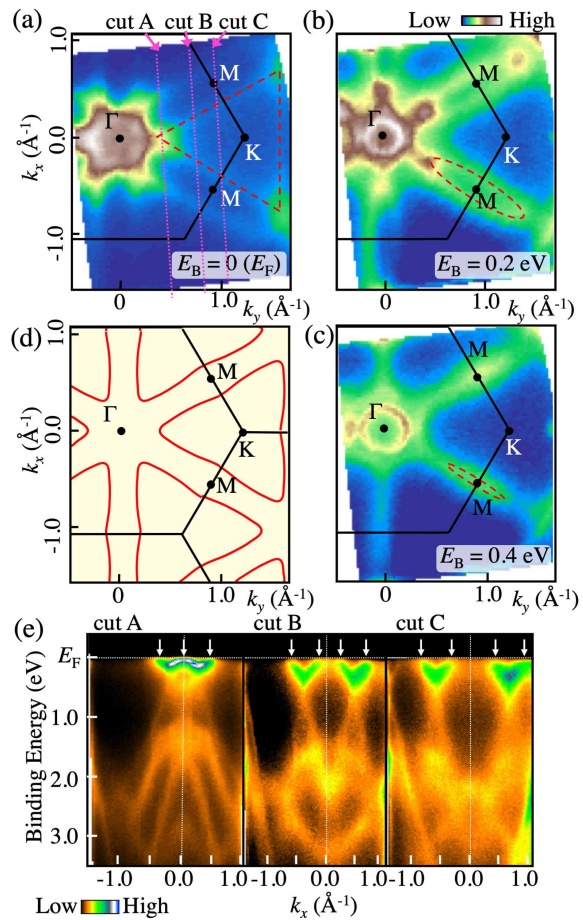}
\caption{(a)-(c) Plots of ARPES intensity at $T$ = 40 K as a function of 2D wave vector, $k_x$ and $k_y$, at three representative energy slices at $E_{\rm B}$ = $E_{\rm F}$, 0.2 eV, and 0.4 eV, respectively. Energy contours were obtained by integrating the intensity within $\pm$ 50 meV with respect to each $E_{\rm B}$'s. (d) Calculated FS obtained from the first-principles band-structure calculations for monolayer 1$T$-VTe$_2$. (e) ARPES-derived band structure along three representative ${\bm k}$ cuts [cuts A-C in (a)] which cross the triangular FS. The systematic evolution of the V-shaped band dispersions from cut A to cut C indicates the holelike nature of triangular FS.}
 \end{center}
\end{figure}

\begin{figure*}
\begin{center}
\includegraphics[width=6.2in]{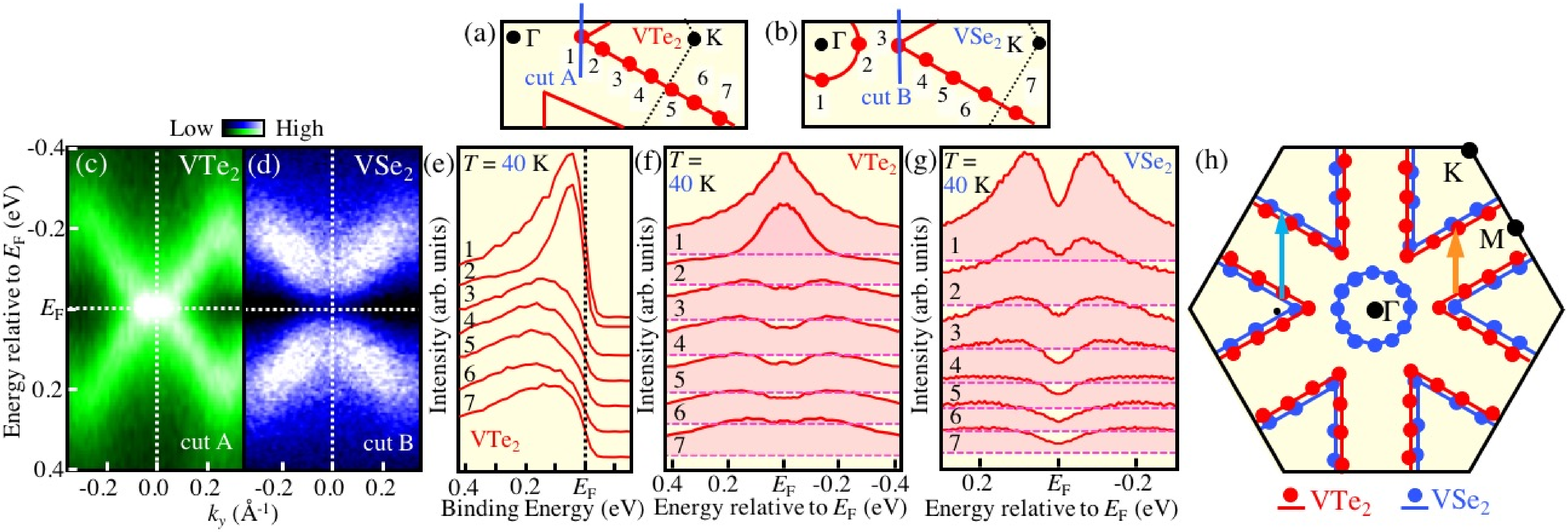}
\caption{(a),(b) Schematic FS together with a ${\bm k}$ cut and ${\bm k}$ points where high-resolution ARPES measurements were performed for monolayer VTe$_2$ and VSe$_2$, respectively. (c),(d) Plots of ARPES intensity as a function of wave vector and $E_{\rm B}$ symmetrized with respect to $E_{\rm F}$, measured at $T$ = 40 K along a cut crossing the corner of triangular pocket (cut A/B) for monolayer VTe$_2$ and VSe$_2$, respectively. (e) EDCs near $E_{\rm F}$ at $T$ = 40 K for monolayer VTe$_2$ measured at various $k_{\rm F}$ points in (a). (f) Same as (e) but symmetrized with respect to $E_{\rm F}$. Zero intensity for each spectrum is indicated by a dashed magenta line to highlight the absolute spectral weight. (g) Same as (f) but for monolayer VSe$_2$ \cite{Umemoto}. (h) Comparison of the FS topology between monolayer VTe$_2$ and VSe$_2$. Red and blue circles correspond to the $k_{\rm F}$ points for monolayer VTe$_2$ and VSe$_2$, respectively. Orange and light blue arrows indicate possible nesting vectors $q$ for monolayer VTe$_2$ and VSe$_2$, respectively.}
 \end{center}
\end{figure*}

Figures 2(a) and 2(b) display the plot of valence-band ARPES intensity for monolayer VTe$_2$ measured at $T$ = 40 K along the $\Gamma$M and $\Gamma$K cuts with $h\nu$ = 56 eV, compared with the corresponding band dispersion obtained by the first-principles band-structure calculations for free-standing monolayer 1$T$-VTe$_2$ with the input of experimental lattice constant $a$. One can see in Fig. 2(a) several energy bands whose dispersion appears to be symmetric with respect to the $\Gamma$ point. A side-by-side comparison of Figs. 2(a) and 2(b) also reveals a good agreement in the overall valence-band structure between the experiment and calculations, demonstrating that the fabricated monolayer VTe$_2$ indeed takes the 1$T$ structure (see section 2 of Supplemental Material for details \cite{SM}). According to the calculations, energy bands lying at the binding energy ($E_{\rm B}$) of 1-6 eV, including the holelike bands which rapidly move toward $E_{\rm F}$ on approaching the $\Gamma$ point, are attributed to the Te 5$p$ orbitals, while the energy band within 0.5 eV of $E_{\rm F}$ with a relatively flat dispersion around the M point is assigned as the V 3$d$ band.
 
To see more clearly the electronic states responsible for the physical properties, we show in Figs. 2(c) and 2(d) the ARPES intensity near $E_{\rm F}$ at $T$ = 40 K measured along the $\Gamma$M and $\Gamma$K cuts, respectively. A detailed spectral analysis by tracing the peak position of EDCs suggests that two topmost Te 5$p$ bands with different band velocities do not reach $E_{\rm F}$, but are topped at 60 meV below $E_{\rm F}$ at the $\Gamma$ point. These bands are degenerate exactly at the $\Gamma$ point, consistent with the calculation in Fig. 2(b). We found that the shallow V 3$d$ band is also located at $\sim$ 60 meV below $E_{\rm F}$ at the $\Gamma$ point and disperses toward higher $E_{\rm B}$ on approaching the M point [Fig. 2(c)], while it crosses $E_{\rm F}$ midway between the $\Gamma$ and K points, accompanied with a sudden drop in the spectral weight [Fig. 2(d)]. While the overall experimental band structure shows a good agreement with the calculated band structure for the 1$T$ phase, we found that some bands near $E_{\rm F}$ in the experiment are renormalized with respect to those in the calculations (for details, see section 2 of Supplemental Material \cite{SM}). It is noted here that we found no evidence for the energy splitting of bands associated with possible exchange splitting due to ferromagnetism, which is further corroborated by our x-ray magnetic circular dichroism measurement at 80 K showing no change in the V $L_{\rm 2,3}$-absorption edge across the magnetic-field reversal \cite{Nakata_inprep}. This suggests the absence of ferromagnetic order in monolayer VTe$_2$.  
At this stage, it is unclear why the ferromagnetism appears in monolayer VSe$_2$ but not in monolayer VTe$_2$, though it is noted that the ferromagnetism in monolayer VSe$_2$ itself is contradictory and is currently a target of fierce debate \cite{Bonilla,Umemoto,Feng}. It is also unknown whether or not the ferromagnetic property is related to the CDW.

To clarify the topology of FS, we have performed ARPES measurements in 2D ${\bm k}$ space. Figures 3(a)-3(c) show the contour maps of ARPES intensity for different $E_{\rm B}$ slices. At $E_{\rm B}$ = $E_{\rm F}$ [Fig. 3(a)], one can recognize a couple of fairly straight intensity patterns around the M point running parallel to the $\Gamma$M direction (red dashed line). This intensity pattern forms a large, almost perfectly triangular-shaped FS enclosing the K point. Remarkably, this FS is well reproduced by the calculations for free-standing 1$T$-VTe$_2$ [Fig. 3(d)] with the input of experimental lattice constant, confirming again the 1$T$ nature of our epitaxial film. Absence of any spurious intensity that could be associated with the band folding with (3 $\times$ 1) periodicity expected from the formation of double zigzag-chain superstructure seen in bulk VTe$_2$ \cite{Ohtani} further corroborates the purely 1$T$ nature of the film (see section 4 of Supplemental Material for details \cite{SM}).

Upon increasing $E_{\rm B}$ to 0.2 eV [Fig. 3(b)], the experimental triangular pattern seen at $E_{\rm B}$ = 0 eV [Fig. 3(a)] transforms into a M-point-centered ellipsoid elongated along the $\Gamma$M direction, which shrinks on further increasing $E_{\rm B}$ to 0.4 eV [Fig. 3(c)]. This indicates that the triangular FS forms a hole pocket, consistent with the calculated band dispersion in Fig. 2(b) in which the V 3$d$ band is located at $\sim$ 1 eV above $E_{\rm F}$ at the K point.
Figure 3(e) shows the ARPES-derived band structure along three representative $k$ cuts (cuts A-C) which cross the triangular FS. On cut A which touches the corner of triangular FS, one can see a couple of V-shaped bands in the vicinity of $E_{\rm F}$.  These two V-shaped bands are gradually separated from each other on going from cut A to cuts B and C, indicating that the triangular FS is holelike.  In Figs. 3(b) and 3(c), one can also identify an intense circular spot at the $\Gamma$ point stemming from the Te 5$p$ bands. We emphasize again that although the proximity of Te 5$p$ bands to $E_{\rm F}$ enhances the intensity at the $\Gamma$ point, these fully occupied bands do not participate in the FS. Therefore, the FS of monolayer VTe$_2$ is solely dictated by the triangular hole pocket at the K point, which greatly simplifies the discussion on the FS topology and nesting, as detailed later. We have estimated the total carrier concentration to be 0.98 $\pm$ 0.08 electrons / unit cell, by evaluating the area of FS with respect to that of whole BZ. This suggests that our monolayer film keeps stoichiometry and no observable charge transfer from the substrate takes place.
 
Now that the FS topology is established, we shall address a key question regarding a possible energy gap opening associated with the occurrence of CDW. We selected a ${\bm k}$ cut passing the corner of triangular FS [blue line in Fig. 4(a)], and show the ARPES intensity at $T$ = 40 K plotted as a function of wave vector and $E_{\rm B}$ symmetrized with respect to $E_{\rm F}$ in Fig. 4(c). One can clearly see a dispersive band reaching $E_{\rm F}$ showing the brightest intensity at $E_{\rm F}$, indicating the absence of an energy gap. This is in sharp contrast to the result of monolayer VSe$_2$ [Figs. 4(b) and 4(d)] that signifies a marked suppression of intensity within $\pm$ 0.1 eV of $E_{\rm F}$ at $T$ = 40 K due to the CDW-gap opening \cite{Umemoto}. To see the low-energy spectral feature in more detail, we have performed high-resolution ARPES measurements along several cuts crossing the FS, and show the EDCs at various $k_{\rm F}$ (Fermi wave vector) points (points 1-7) covering the whole straight segment of the triangular FS in the first BZ in Fig. 4(e). The corresponding symmetrized EDCs in Fig. 4(f) show a single peak at points 1 and 2 located around the corner of FS, while the EDCs at points 3-7 exhibit a weak dip structure at $E_{\rm F}$ indicative of a large residual spectral weight at $E_{\rm F}$. We attribute this spectral-weight suppression as the pseudogap, but not the CDW gap, since the spectral behavior resembles that of monolayer VSe$_2$ at room temperature (well above $T_{\rm CDW}$ $\sim$ 140 K), which shows the coexistence of pseudogap and metallic Fermi-arc states \cite{Umemoto}. Also, the pseudogap of VTe$_2$ persists over a wide temperature range (10 - 300 K), similarly to VSe$_2$. It is noted that the pseudogap is unlikely to be due to some extrinsic effects such as the sample/surface quality and/or the experimental conditions (e.g. photoionization cross-section and light polarization), but is an intrinsic feature of monolayer VTe$_2$. The pseudogap may be explained in terms of the CDW fluctuations and/or the electron-phonon coupling associated with the CDW (see also section 3 of Supplemental Material \cite{SM}).  The metallic state revealed in Figs. 4(c) and 4(e) in VTe$_2$ is obviously different from the fully gapped insulating state below $T_{\rm CDW}$, as visible from the strong spectral-weight suppression around $E_{\rm F}$ over the entire FS as seen in Fig. 4(g). We thus conclude that the CDW is suppressed in monolayer VTe$_2$.

A key to understand such a contrasting behavior may lie on the difference in the FS topology between the two V-dichalcogenide monolayers. Figure 4(h) directly compares the FS obtained from ARPES measurements of monolayer VTe$_2$ and VSe$_2$ \cite{Umemoto}. One can immediately recognize that both monolayers show a similar triangular pocket at the K point whereas a circular hole pocket exists only in VSe$_2$. Extra hole carriers at the $\Gamma$-centered pocket in VSe$_2$ resides on the K-centered pocket in VTe$_2$, as seen from a larger triangular pocket in VTe$_2$. This is reasonable since the Se and Te atoms are isovalent and the total FS area should be the same. The expansion of triangular pocket would widen the straight segment of the FS in VTe$_2$. Assuming that the nesting vector $q$ is parallel to the $\Gamma$M direction \cite{Umemoto}, this would lead to an enhancement of electronic susceptibility in VTe$_2$. Thus, one would naively expect that the CDW in VTe$_2$ is more stable than that in VSe$_2$ according merely to the FS-nesting picture. However, this is not the case since the CDW appears to be suppressed in VTe$_2$. Thus, one cannot sufficiently describe the CDW of monolayer VSe$_2$ or VTe$_2$ simply in terms of the energy gain around $E_{\rm F}$ in the electronic system, which suggests the importance of considering the electron-phonon coupling \cite{Calandra}. While detailed discussion on the relevance of electron-phonon coupling requires sophisticated first-principles band-structure calculations, we point out here a possibility that such an electron-phonon coupling could be linked to the electronic states via the commensurability of the nesting. We found that the nesting vector along the $\Gamma$M direction in VSe$_2$ is commensurate to the lattice ( 1/4 $\bf{G}$ where $\bf{G}$ is the reciprocal lattice vector) \cite{Umemoto}, while that in VTe$_2$ is incommensurate ( 1/4.6 $\bf{G}$). If such commensurability enhances the electron-phonon coupling at the corresponding lattice periodicity, it may stabilize the CDW. However, this explanation is still speculative, requiring further experimental and theoretical studies to firmly pin down the CDW origin.

Our band calculations of monolayer VTe$_2$ show that a small (1.5 $\%$) change in the in-plane lattice constant is sufficient to control the emergence/absence of a small hole pocket at $\Gamma$ and the concomitant change in the FS-nesting condition at the triangular pocket. Such sensitivity of the FS topology to the lattice parameters is essential due to the fact that the narrow V 3$d$ band is located in the vicinity of $E_{\rm F}$. Therefore, we expect that small perturbations such as lattice strain and carrier doping would easily trigger the change in the FS topology and the nesting vector, leading to the modulation of CDW properties. In this regard, the reported differences in the FS topology around the $\Gamma$ point in VSe$_2$, which may link to diverse periodic lattice distortions \cite{Umemoto, Chen, Feng, Duvjir}, may be interpreted in terms of a reflection of the high sensitivity of the CDW characteristics to the strain and carrier balance. The present result suggests an importance of precisely controlling the lattice strain and carrier concentration for manipulating the CDW. Such band engineering would be a main target of future studies.
 
In conclusion, we have performed an ARPES study on monolayer 1$T$-VTe$_2$ grown on bilayer graphene by MBE. We found a large triangular FS at the K point that satisfies a nearly perfect nesting condition, whereas the CDW is suppressed as highlighted by the observation of $E_{\rm F}$-crossing of bands at low temperature, in contrast to monolayer VSe$_2$ that exhibits a well-defined CDW characterized by the fully gapped insulating state.  The present result opens a pathway toward controlling novel physical properties of 2D TMDs through the band engineering.

\begin{acknowledgments}
We thank Y. Umemoto, K. Horiba, and H. Kumigashira for their assistance in the ARPES experiments. This work was supported by the MEXT of Japan (Innovative Area ``Topological Materials Science'' JP15H05853), JST-PREST (No. JPMJPR18L7), JST-CREST (No. JPMJCR18T1), JSPS KAKENHI Grants (No. JP18K18986, JP18H01821, JP18H01160, JP17H04847 and JP17H01139), KEK-PF (Proposal No. 2018S2-001), Science research projects from Murata Science Foundation, World Premier International Research Center, Advanced Institute for Materials Research. Y. N. acknowledges support from GP-Spin at Tohoku University.
\end{acknowledgments}

\bibliographystyle{prsty}

\end{document}